\documentclass[useAMS,usenatbib]{mn2e}
\usepackage{graphicx}
\usepackage{epstopdf}
\usepackage{amsmath}
\usepackage{amssymb}
\title[Optical/UV emission of TDEs]{The origin of the optical/ultraviolet emission of optical/ultraviolet tidal disruption events}
\author[Bu, Qiao, Yang \& Liu]{De-Fu Bu$^1$\thanks{E-mail: dfbu@shao.ac.cn}, Erlin Qiao$^{2,3}$\thanks{E-mail: qiaoel@nao.cas.cn}, Xiao-Hong Yang$^4$\thanks{E-mail: yangxh@cqu.edu.cn} and Jifeng Liu$^{2,3}$ \\
$^{1}$Shanghai Astronomical Observatory, Chinese Academy of Sciences, 80 Nandan Road, Shanghai 200030, China \\
$^{2}$Key Laboratory of Space Astronomy and Technology, National Astronomical Observatory, \\ Chinese Academy of Sciences, Beijing 100012, China \\
$^3$School of Astronomy and Space Sciences, University of Chinese Academy of Sciences, 19A Yuquan Road, Beijing 100049, China \\
$^4$Department of Physics, Chongqing University, Chongqing 400044, China}
\begin{document}

\pagerange{\pageref{firstpage}--\pageref{lastpage}} \pubyear{2002}

\maketitle

\label{firstpage}

\begin{abstract}
One of the most prominent problems of optical/ultraviolet (UV) tidal disruption events (TDEs) is the origin of their optical/UV emission. It has been proposed that the soft X-rays produced by the stellar debris accretion disk can be reprocessed into optical/UV photons by a surrounding optically thick envelope or outflow. However, there is still no detailed models for this mechanism. In this paper, by performing hydrodynamic simulations with radiative transfer, we calculate the optical/UV emission of the circularized stellar debris accretion flow/outflow system. We find that the optical/UV photons can be generated by reprocessing the emission of the accretion flow in the optically thick outflows. The model can well interpret the observed emission properties of optical/UV TDEs, including the emission radius, the radiation temperature and the luminosity, as well as the evolution of these quantities with time, providing a strong theoretical basis for understanding the origin of optical/UV TDEs.
\end{abstract}


\section{Introduction}
A star can be tidally disrupted as it approaches close enough to the supermassive black hole in the center of a galaxy, triggering the so-called tidal disruption events (Rees 1988; Evans \& Kochanek 1989; Ulmer 1999). After disruption, the unbound debris can escape; while the bound debris will fall back. The collisions between fall back debris streams can induce shocks, which dissipate the orbital energy of debris. It is proposed that this process may `circularize' the debris to form circularized accretion flow. However, the efficiency of circularization by shock is still under debates (Kochanek 1994; Guillochon \& Ramirez-Ruiz 2015; Shiokawa et al. 2015; Dai et al. 2015; Bonnerot et al. 2016; Hayasaki et al. 2016;Bonnerot et al. 2017; Lu \& Bonnerot 2020; Bonnerot \& Lu 2020; Bonnerot \& Stone 2021; Rossi et al. 2021).

TDEs were first identified in soft X-ray band with the data of ROSAT all-sky survey in 1990-1991 (Bade, Komossa \& Dahlem 1996; Komossa \& Bade 1999; Komossa \& Greiner 1999; Grupe et al. 1999; Greiner et al. 2000; Komossa 2015), which are roughly consistent with the prediction of emission of a circularized accretion disk around a supermassive  black hole (with black hole mass $\sim 10^{6-7}$ times solar mass). However, currently, a growing number of TDEs are discovered in optical/UV band (Gezari et al. 2008; van Velzen et al. 2011; Arcavi et al. 2014; Gezari et al. 2012; Chornock et al. 2014; Holoien et al. 2014; Holoien et al. 2016b; Holoien et al. 2016a; Blagorodnova et al. 2017), and the number of optical/UV discovered TDEs has accounted for more than two thirds of the totally discovered TDEs (see van Velzen et al. 2020 and Gezari 2021 for review).

One of the most prominent problems of optical/UV TDEs is the origin of their emission. The inferred radius of optical/UV radiation for Optical/UV TDEs is $\sim 10^{14-16}$ cm (Hung et al. 2017; van Velzen et al. 2020; Gezari 2021), which is orders of magnitude larger than the predicted size of the accretion disk of a few times $10^{13}$ cm (assuming a solar type star disrupted by a black hole with $10^6-10^7$ solar mass). There are two debated origins of optical/UV emission. In the first scenario, the emission is powered by the shocks between the debris streams as they collide (Piran et al. 2015; Jiang et al. 2016; Steinberg \& Stone 2022). In this scenario, the shock/emission radii can be consistent with that inferred from observations. In the second scenario, the soft X-ray/EUV emission generated by the debirs accretion onto the black hole is reprocessed into optical/NUV bands by an extended elliptical disk (Liu et al. 2017, 2021; Wevers et al. 2022) or an envelope
(Loeb \& Ulmer 1997; Coughlin \& Begelman 2014; Roth et al. 2016) or an outflow (Strubbe \& Quataert 2009; Lodato \& Rossi 2011; Metzger \& Stone 2016; Parkinson et al. 2022) surrounding the central debris disk. The `reprocessing' outflows may have two origins. First, if the debris can be quickly `circularized', a super-Eddington accretion flow will form. An optically thick outflow can be driven by radiation pressure from a super-Eddington accretion flow
(Dai et al. 2018; Curd \& Narayan 2019; Thomsen et al. 2022). Second, an outflow can also be driven in the shock process (Jiang et al. 2016; Lu \& Bonnerot 2020). In the reprocessing scenario, the location of optical/UV emission is also consistent with observations.

For the optical/UV TDEs, there are four main observational features (Hung et al. 2017; van Velzen et al. 2020; Gezari 2021). First, the emission radius is $\sim 10^{14-16}$ cm from the central black hole. Second, the radiation temperature is in the range of $10^4-5 \times 10^4$K. Third, the peak luminosity is in the range of $10^{43}-10^{45}$ erg/s. Fourth, the emission radius decreases slightly with time, and the radiation temperature is nearly constant with time. Both the shock and reprocessing models have tried to interpret these observational features. These two models can roughly interpret the emission radius and radiation temperature of optical/UV TDEs. However, the time evolution of the emission radius and radiation temperature with time has not been investigated by these two models. In this paper, by performing radiation hydrodynamic simulations, we study whether the four observational features of optical/UV TDEs can be interpreted by the circularized super-Eddington accretion flow/outflow system.

This paper is structured as follows. In Section 2, we present the detailed equations, numerical settings of the simulations. In Section 3, we present our results. We summarize and discuss our results in Section 4.

\section{Numerical method }
We have performed two-dimensional axisymetric hydrodynamic simulations with radiative transfer. We have developed the radiative transfer module for simulations (Yang et al. 2014; Yang \& Yuan 2012). We incorporate our radiative transfer module to the PLUTO code (Mignone et al. 2007) to perform our simulations. We adopt the spherical coordinates ($r$, $\theta$, $\phi$). The equations are as follows,
\begin{equation}
 \frac{d\rho}{d t} + \rho \nabla\cdot {\bf v} = 0,
\end{equation}
\begin{equation}
\rho\frac{d {\bf v}}{dt}=-\nabla p-\rho \nabla\psi+ \nabla \cdot {\bf T} + {\bf f}_{\rm rad},
\end{equation}
\begin{equation}
\rho\frac{d}{dt}\Big(\frac{e}{\rho}\Big)=-p\nabla\cdot {\bf v} -4\pi \kappa B +c \kappa E_{\rm rad} + Q_{\rm vis}
\end{equation}
\begin{equation}
\rho\frac{d}{dt}\Big(\frac{E_{\rm rad}}{\rho}\Big)= -\nabla \cdot {\bf F}_0 -\nabla {\bf v} : {\bf P}_0 + 4\pi \kappa B - c \kappa E_{\rm rad}
\end{equation}
Here, $\rho$, ${\bf v}$, $p$, $e$ and $E_{\rm rad}$ are gas density, velocity, gas pressure, gas internal energy density and radiation energy density, respectively. We employ an adiabatic equation of state $p=(\gamma-1)e$, with $\gamma=5/3$. $\psi$ is the gravitational potential of the black hole. We adopt the pseudo-Newtonian potential, $\psi=-GM_{\rm BH}/(r-R_{s})$, with $G$, $M_{\rm BH}$ and $R_s$ being the gravitational constant, the black hole mass and the Schwarzschild radius, respectively. ${\bf T}$ is the anomalous stress tensor, which is responsible for angular momentum transfer. $Q_{\rm vis}$ is the viscous heating term. ${\bf F}_0$ is the radiation flux. ${\bf P}_0$ is the radiation pressure tensor. $\kappa$ is the absorption opacity. ${\bf f}_{\rm rad}$ is the radiation force. $B$ is the blackbody intensity. $c$ is the speed of light.

The absorption opacity $\kappa=\kappa_{ff}+\kappa_{bf}$, with $\kappa_{ff}$ and $\kappa_{bf}$ being the free-free absorption and bound-free absorption, respectively. The free-free absorption $\kappa_{ff}=1.7\times 10^{-25}T^{-7/2}(\frac{\rho}{m_p})^2 \ {\rm cm}^{-1}$ and the bound-free absoprtion $\kappa_{bf}=4.8\times 10^{-24}T^{-7/2}(\frac{\rho}{m_p})^2(\frac{Z}{Z_\odot}) \ {\rm cm}^{-1}$
(Hayashi et al. 1962; Rybicki \& Lightman 1979), where $Z$, $Z_\odot$ and $T$ are metallicity, solar metallicity and gas temperature, respectively. We assume that $Z=Z_\odot$.  The radiation force ${\bf f}_{\rm rad}=\frac{\chi}{c} {\bf F}_0$, $\chi$ is the total opacity. $\chi=\kappa+\rho\kappa_{es}$, where $\kappa_{es}$ is the electron scattering opacity. We assume the gas to have solar chemical abundances. Therefore, $\kappa_{es}= 0.34 {\rm cm^2 g^{-1}}$. We use flux-limited diffusion approximation (Levermore \& Pomraning 1981) to bridge the radiation flux and the radiation energy density, ${\bf F}_0=-\frac{c\lambda}{\chi}\nabla E_{\rm rad}$, with $\lambda$ being the flux limiter. $\lambda=\frac{2+\Re}{6+3\Re+\Re^2}$, with $\Re=|\nabla E_{\rm rad}|/\chi E_{\rm rad}$. The radiation pressure tensor ${\bf P}_0={\bf f} E_{\rm rad}$, with ${\bf f}$ being the Eddington tensor. The Eddington tensor ${\bf f}=\frac{1}{2}(1-f){\bf I}+\frac{1}{2}(3f-1){\bf n n}$, with $f=\lambda+\lambda^2\Re^2$ and ${\bf n}=\nabla E_{\rm rad}/|\nabla E_{\rm rad}|$.

We assume that only the $r\phi$ component of the viscous stress tensor is nonzero, $T_{r\phi}=\eta r \frac{\partial}{\partial r}\frac{v_\phi}{r}$, with $\eta$ being the dynamical viscosity coefficient. The viscous heating term is $Q_{\rm vis}=\eta (r\frac{\partial}{\partial r}(\frac{v_\phi}{r}))^2$. As down by Ohsuga et al. (2005), we assume that $\eta=\alpha \frac{p+\lambda E_{\rm rad}}{v_\phi/r}$. In this work, we set the viscous parameter $\alpha=0.1$.

\begin{figure*}
\begin{center}
\includegraphics[scale=0.43]{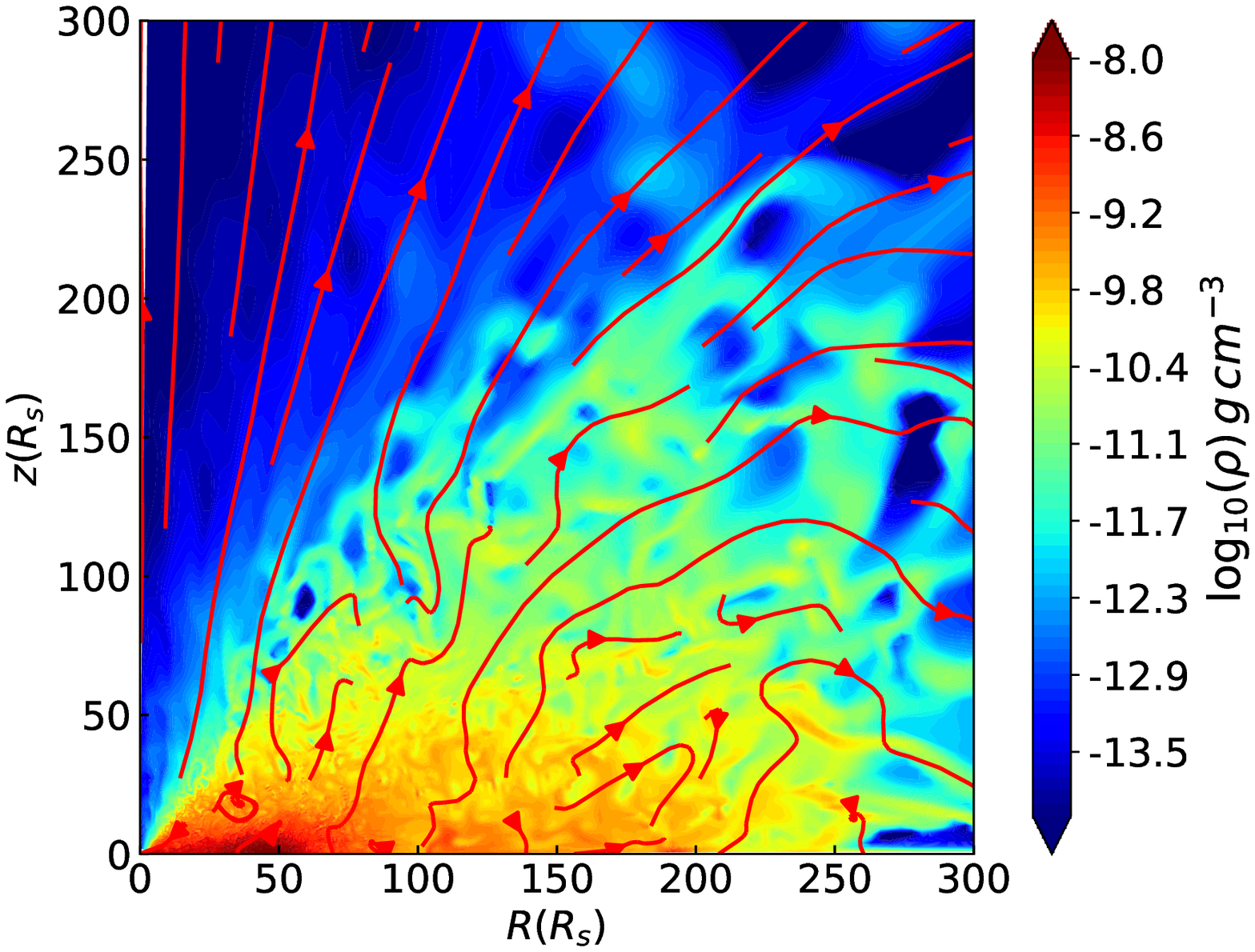}\hspace*{0.1cm}
\includegraphics[scale=0.43]{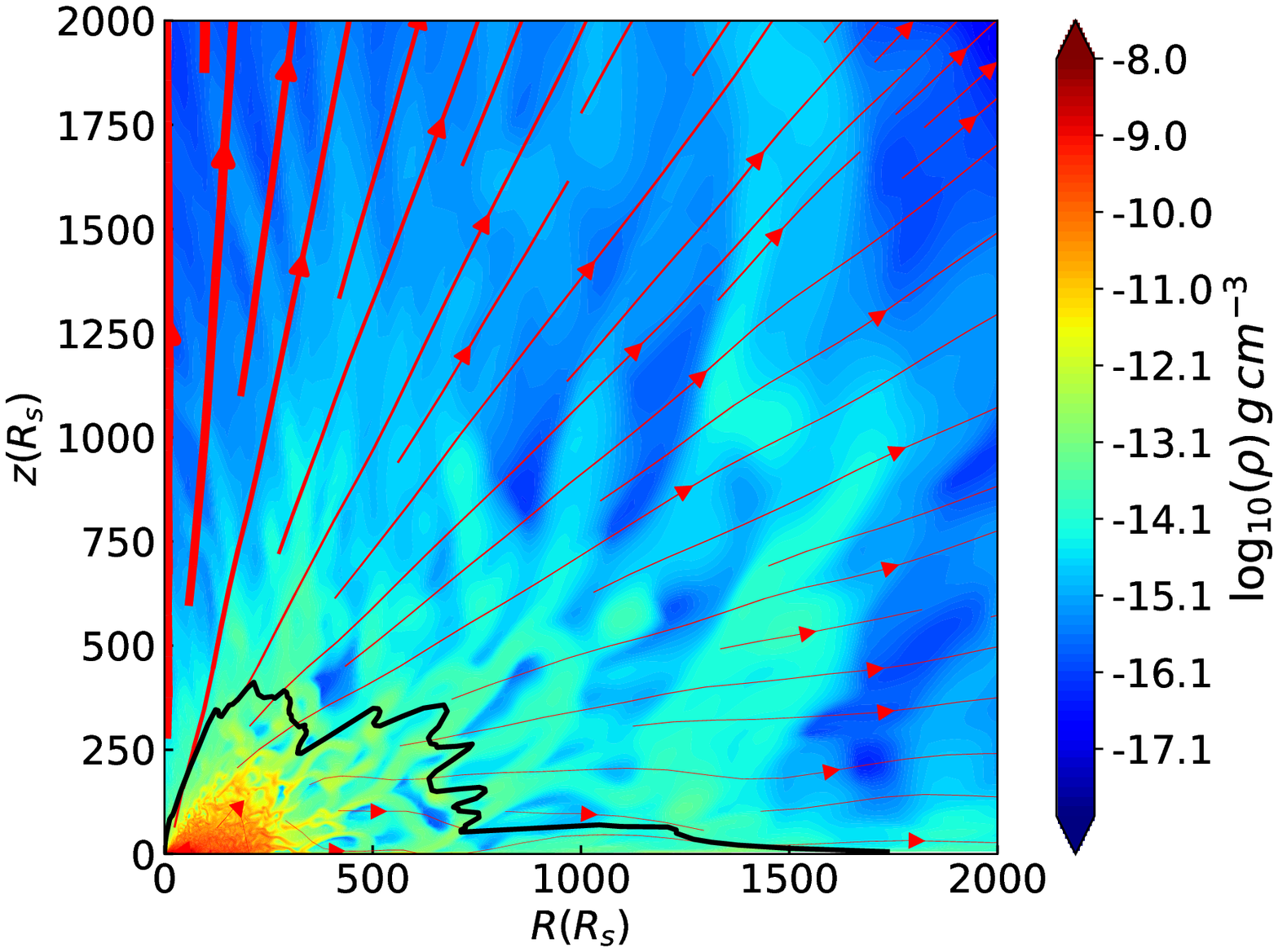}\hspace*{0.7cm}\\
\includegraphics[scale=0.43]{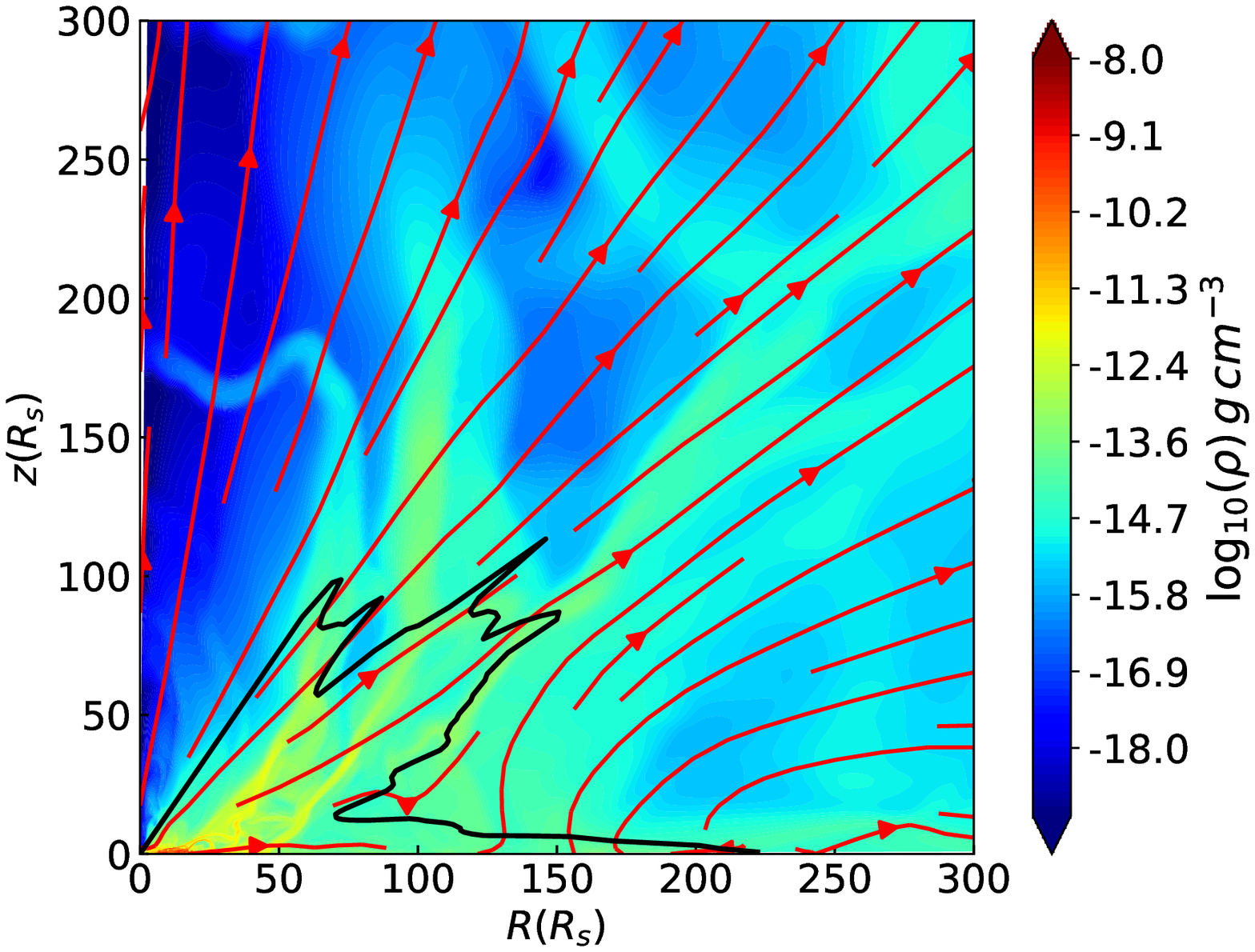}\hspace*{0.1cm}
\includegraphics[scale=0.43]{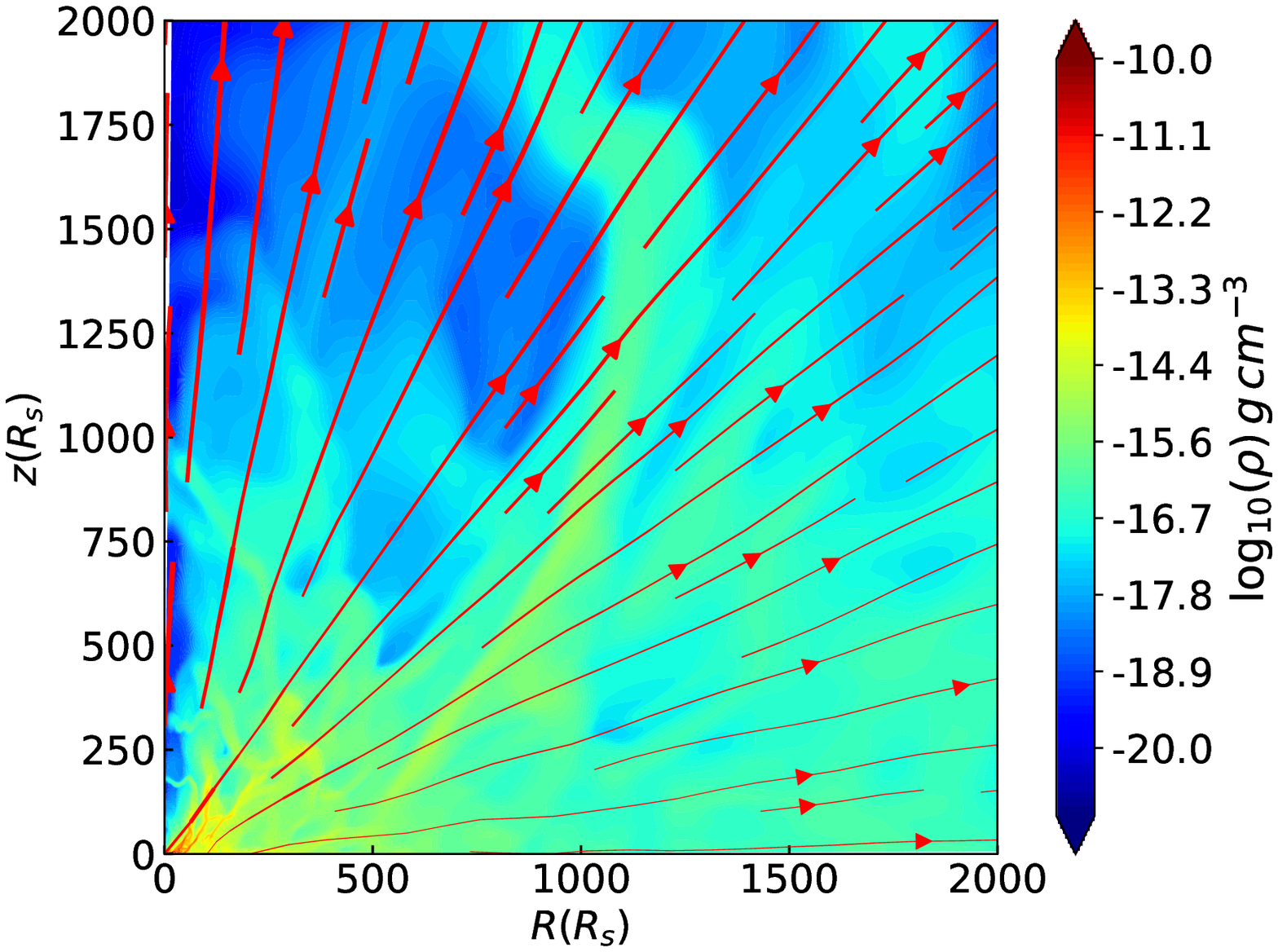}\hspace*{0.7cm}\\
\hspace*{0.5cm} \caption{Snapshots for gas density (color scale) with fluid velocity (streamlines). The upper two panels are for model M6 at $t=15.9$ days ($t$ is measured from the beginning of the simulation). The bottom two panels are for model M7 at $t=95.7$ days. The left two panels show a zoom-in small domain. The right two panels show a much larger domain. The thick black lines in the upper-right and bottom-left panels are the electron scattering photosphere. \label{fig:Dcolor}}
\end{center}
\end{figure*}

When a star moving on a parabolic trajectory towards the black hole, its mechanical energy (kinetic energy and gravitational energy) is zero. In this paper, we assume that the disrupted star moves on a parabolic trajectory and its pericenter is equal to the tidal disruption radius $R_T$ (Hills 1975). We assume that when the disrupted stellar debris falls back, it can be very quickly circularized to form an accretion flow. Due to angular momentum conservation, the circularized accretion flow forms at the circularization radius $R_C$, which is two times the disruption radius $R_T$. We also assume that the disrupted star has radius $R_\ast=R_\odot$ and mass $M_\ast=M_\odot$, with $R_\odot$ and $M_\odot$ being solar radius and solar mass, respectively.

We have two simulations. In simulation M7, we set the black hole mass $M_{\rm BH} = 10^7 M_\odot$ (with $M_\odot$ being solar mass). In simulation M6, the black hole mass $M_{\rm BH} = 10^6M_\odot$. The tidal radius for a $10^6M_\odot$ black hole is at $R_T=47/2R_s$; for a $10^7M_\odot$ black hole, $R_T=5R_s$. After the circularization of the fallback debris, an circularized accretion disk forms at $R_C=2R_T$. In model M7, $R_C=2R_T=3 \times 10^{13} {\rm cm} = 10 R_s$. In model M6, $R_C=2R_T=1.4\times 10^{13} {\rm cm} = 47 R_s$. When a star moving on a parabolic trajectory towards the black hole is disrupted, the stellar debris can have the distribution of the mass with respect to specific binding energy roughly flat. In this case, the bound stellar debris falls back with a fallback rate decreasing with time $\dot M_{fb} \propto t^{-5/3}$ (Rees 1988). Motivated by this theoretical prediction, we inject gas into our computational domain around $R_c$, with the mass injection rate set to equal the gas fall back rate $\dot M_{inject} = \dot M_{fb} = \frac{1}{3}(M_\ast/t_{fb})(1+t/t_{fb})^{-5/3}$, with $t_{fb}$ being the debris fallback timescale. The fallback timescale $t_{fb} \approx 40 \ {\rm days} \ (\frac{M_{\rm BH}}{10^6M_\odot})^{1/2} (\frac{M_\ast}{M_\odot})^{-1}(\frac{R_\ast}{R_\odot})^{3/2} $. For the black hole mass ($10^6 M_\odot$ and $10^7 M_\odot$) in this paper, the peak injection rate is super-Eddington. In model M7, we inject gas in the region $8R_s \leq r \leq 12 R_s$ and $\frac{9}{20}\pi \leq \theta \leq \pi/2$. In model M6, gas is injected in the region $45R_s \leq r \leq 49 R_s$ and $\frac{9}{20}\pi \leq \theta \leq \pi/2$.  In our two models, we assume that the fallback debris can be quickly circularized. Therefore, we assume that the rotational velocity of the injected gas equals to the local Keplerican velocity. We also assume that the internal energy of the fallback debris is not large compared to the gravitational energy. In both models, we assume the internal energy of the injected gas equals $1\%$ of the local gravitational energy. The velocities in $r$ and $\theta$ direction of the injected gas are assumed to be 0.

In both models, computational domain in radial direction is $2R_s \leq r \leq 10^5 R_s$. In the $\theta$ direction, the computational domain is $0 \leq \theta \leq \pi/2$. In $r$ direction, we have 768 grids. In order to well resolve the accretion flow, the grids are not uniformlly spaced, with grids size much smaller close to the center. In the region $2R_s \leq  r \leq 3R_s$, we have 16 uniformly spaced grids. We have 240 logarithm grids in the region $3R_s \leq r \leq 50R_s$. In the region $50R_s \leq r \leq 200R_s$, we have 128 logarithm grids. Finally, in the outer most region $200R_s \leq  r \leq 10^5R_s$, we have 384 logarithm grids. In $\theta$ direction, we have 128 non-uniform spaced grids, with grids size much smaller close to the midplane. At the inner and outer radial boundary, we employ outflow boundary conditions. At the $\theta=0$ boundary, we assume the axisymmetric boundary condition. At the $\theta=\pi/2$ boundary, we employ the reflecting boundary condition.

\section{Results}

\begin{figure*}
\begin{center}
\includegraphics[width=0.5\textwidth]{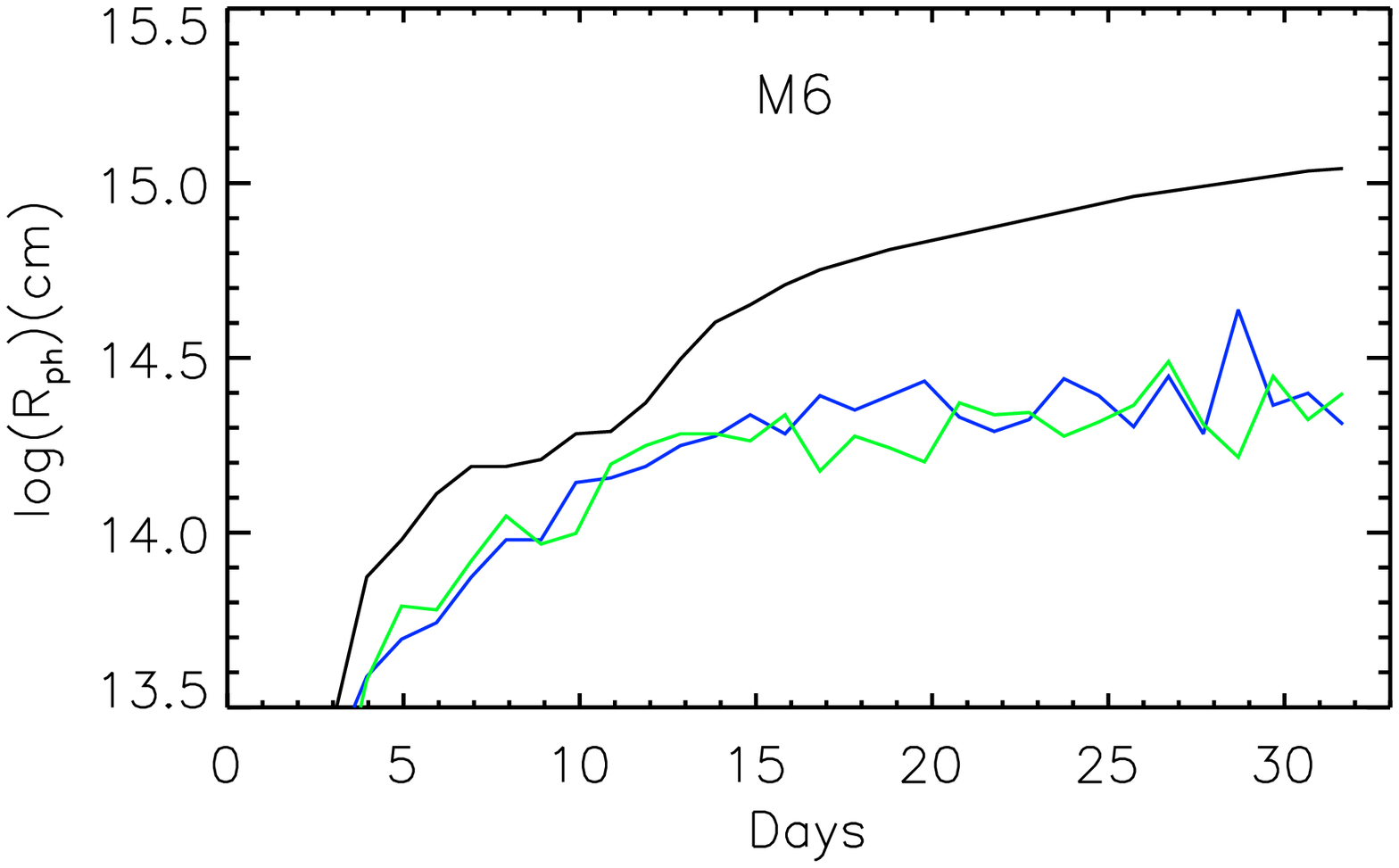}
\includegraphics[width=0.5\textwidth]{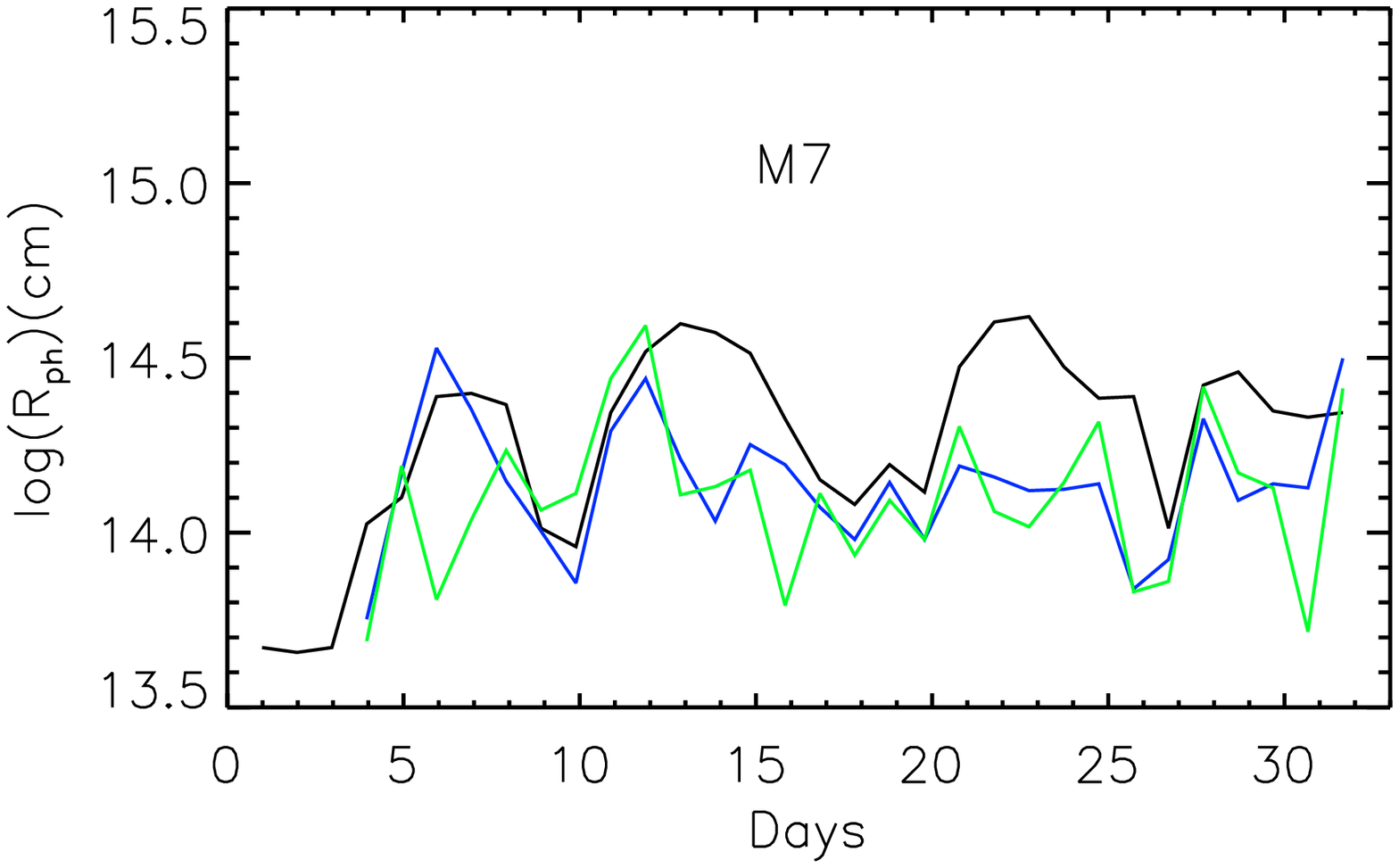}
\includegraphics[width=0.5\textwidth]{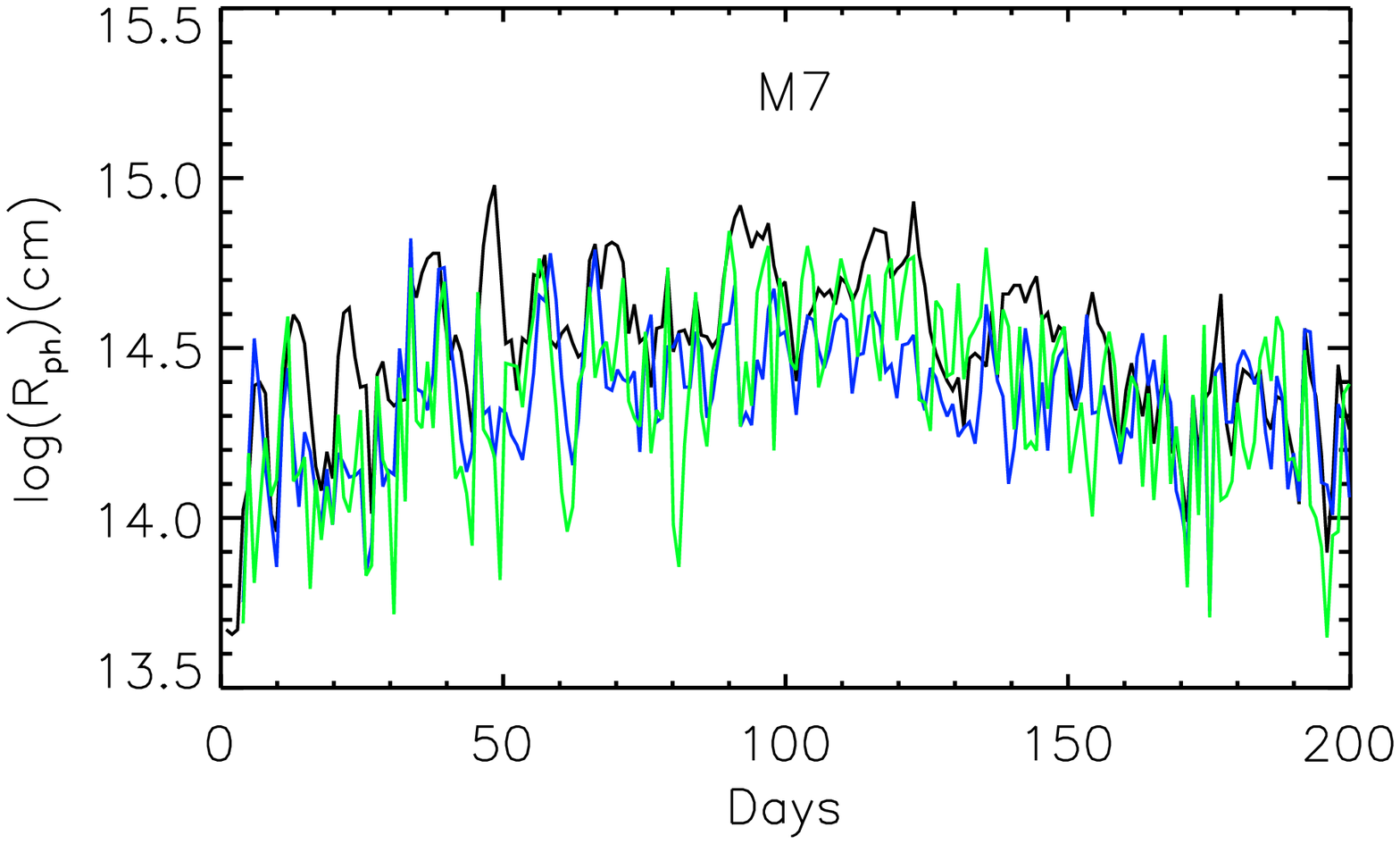}\\
\caption{Time evolution of the electron scattering photosphere along three viewing angles of $60^\circ$ (green line), $75^\circ$ (blue line) and $90^\circ$ (black line). The upper panel is for model M6. The middle and bottom panels are for model M7. The purpose of plotting the middle panel is to clearly illustrate the initial increasing phase of the photosphere radius.}
\label{fig:Rsphere}
\end{center}
\end{figure*}

\begin{figure*}
\begin{center}
\includegraphics[scale=0.43]{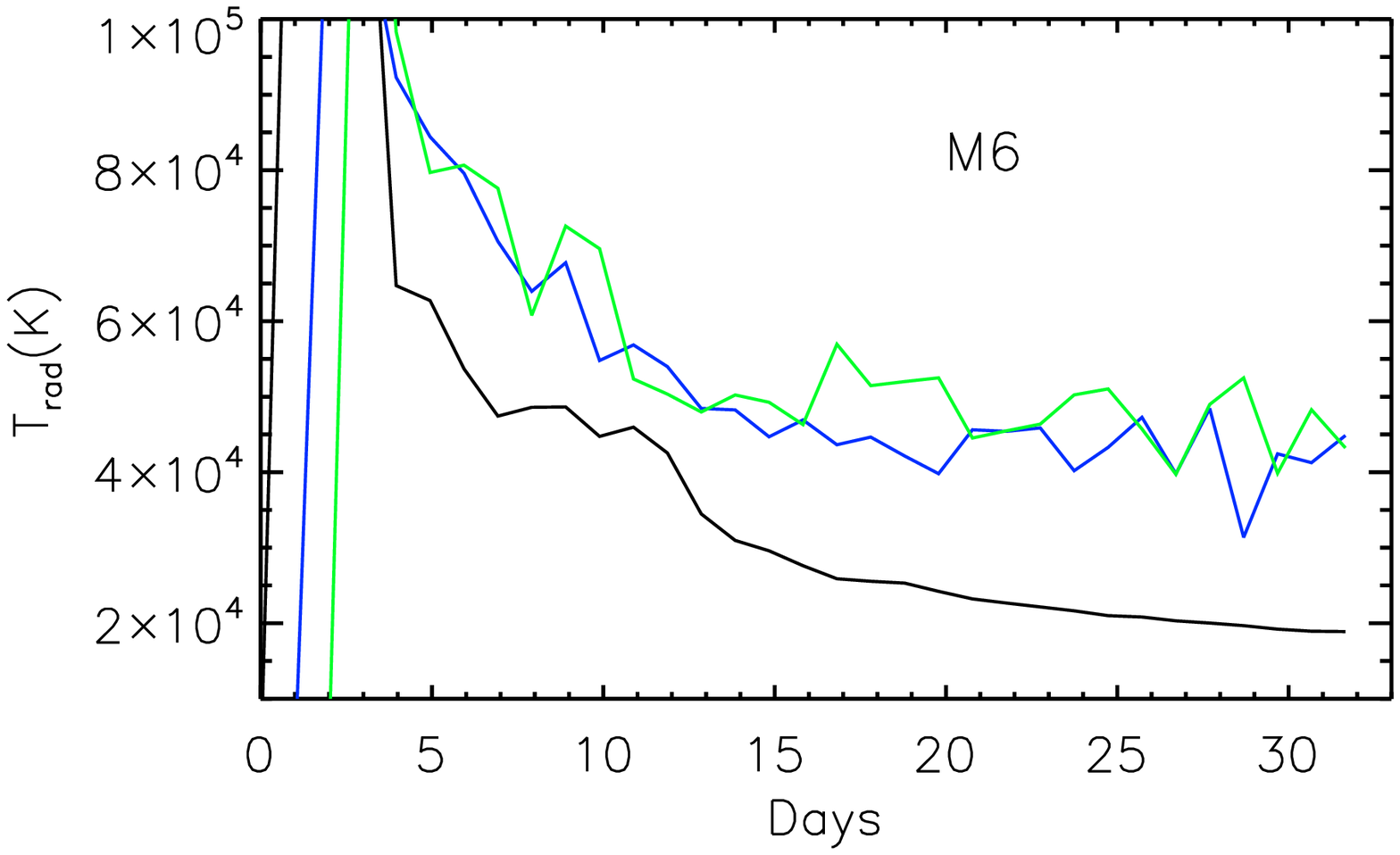}\hspace*{0.1cm}
\includegraphics[scale=0.43]{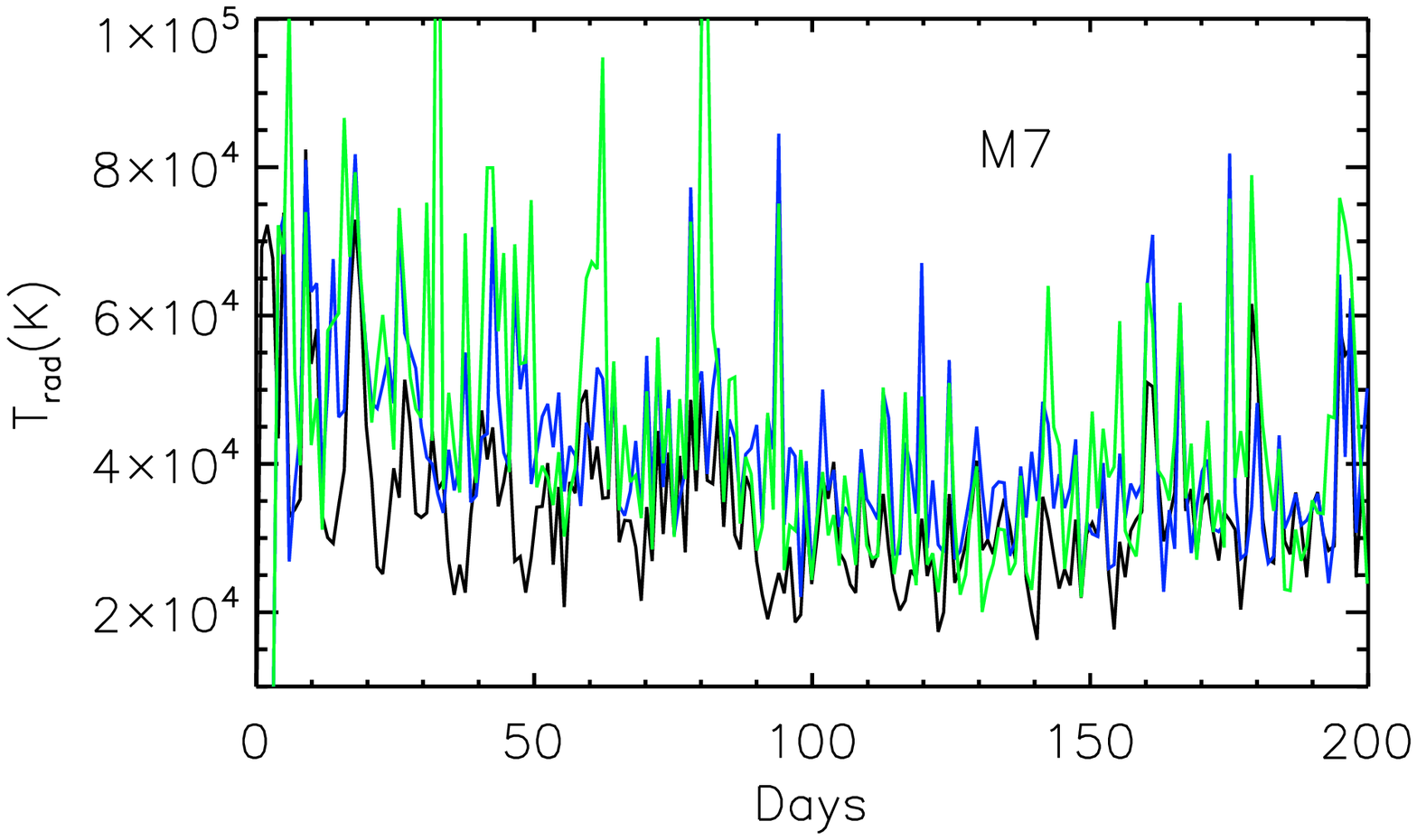}\hspace*{0.7cm}\\
\hspace*{0.5cm} \caption{Time evolution of the radiation temperature at photosphere along three viewing angles of $60^\circ$ (green line), $75^\circ$ (blue line) and $90^\circ$ (black line). The left panel is for model M6. The right panel is for model M7. \label{fig:Trad}}
\end{center}
\end{figure*}

\begin{figure*}
\begin{center}
\includegraphics[scale=0.43]{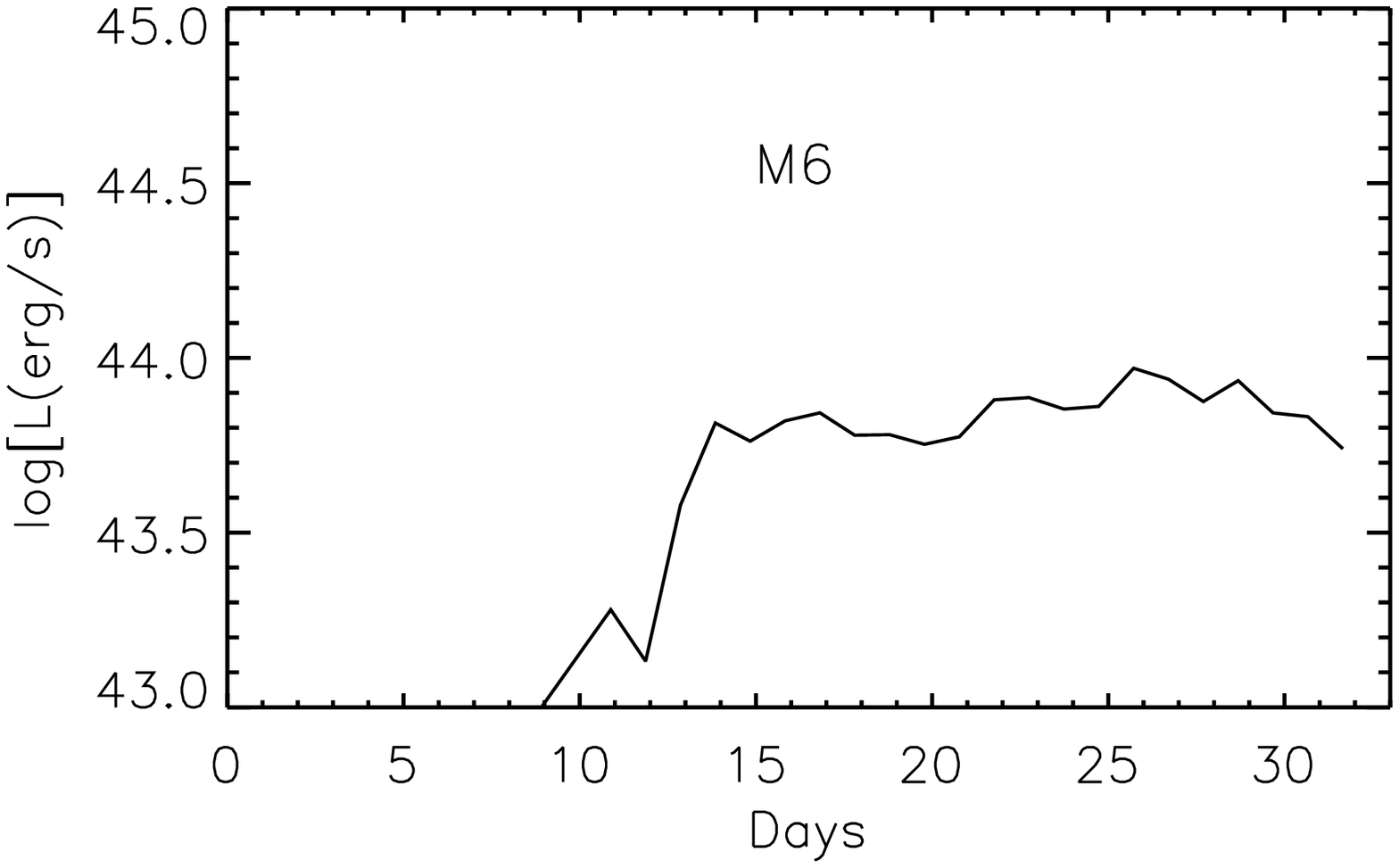}\hspace*{0.1cm}
\includegraphics[scale=0.43]{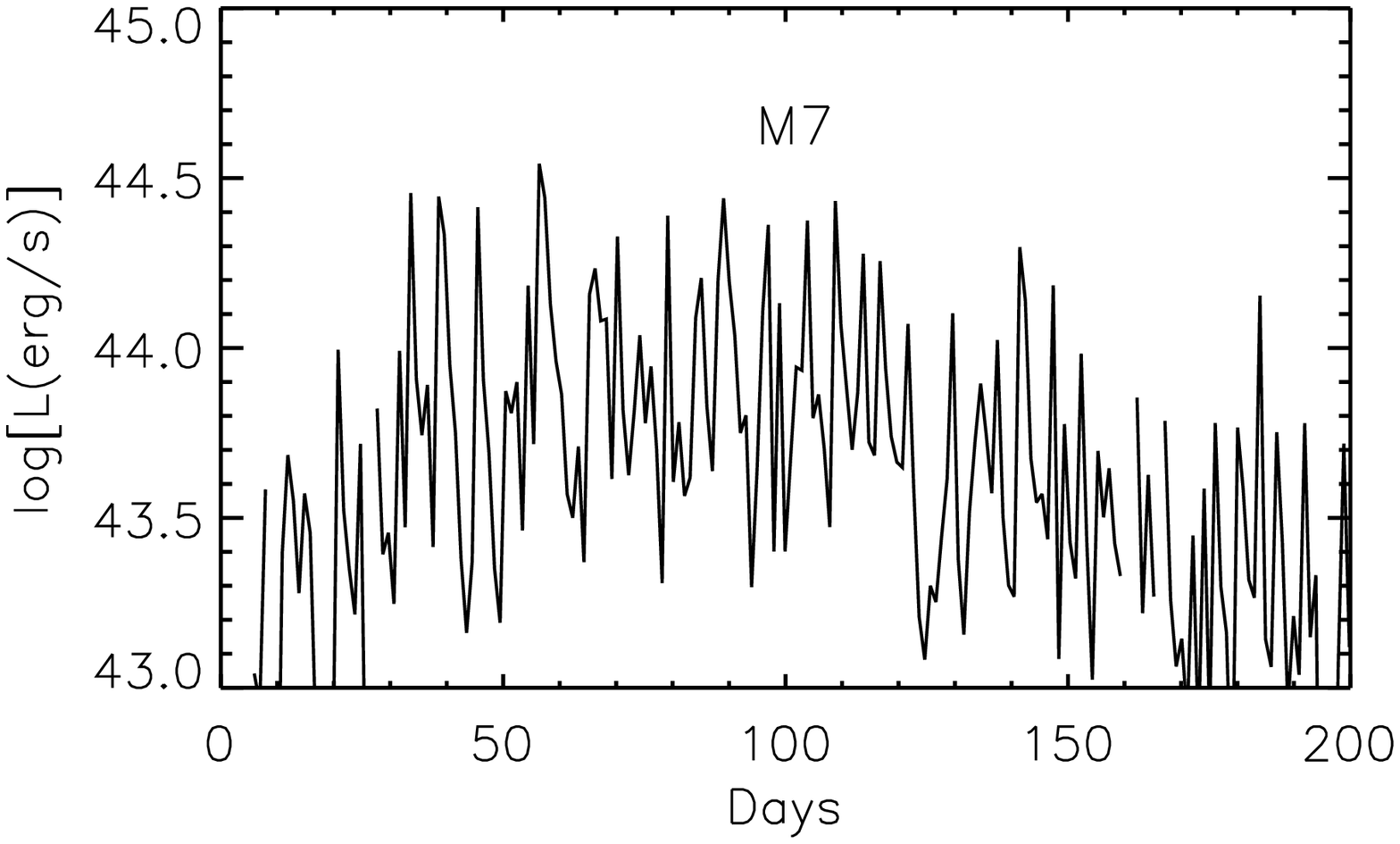}\hspace*{0.7cm}\\
\hspace*{0.5cm} \caption{Time evolution of the optical/UV luminosity emitted at the electron scattering photosphere in the temperature range of $10^4$ K- $5 \times 10^4$K. The left panel is for model M6. The right panel is for model M7. \label{fig:Lum}}
\end{center}
\end{figure*}
Figure \ref{fig:Dcolor} shows the snapshot of gas density and velocity in our two models. In both models, an optically thick outflow can be driven. The outflow can go to significantly larger distance than the gas injection radius. Generally, the density of outflow is highest at the midplane. It decreases towards the gas rotational axis. An electron scattering photosphere forms at viewing angles larger than a critical angle $\theta_{cr}$. Following Dai et al. (2018) and Curd \& Narayan (2019), we define the photosphere to be the location where the electron scattering optical depth is 1.
In order to calculate the electron scattering photosphere along a viewing angle $\theta$, we calculate the electron scattering optical depth from outer radial boundary inwards $\tau (\theta, r)=\int_{10^5R_s}^r \rho \kappa_{es} dr'$. The radial location corresponding to $\tau (\theta, r)=1$ is defined as the photosphere along viewing angle $\theta$. The values of $\theta_{cr}$ are $\sim 10^\circ$ and $\sim 45^\circ$ for models M6 and M7, respectively. For viewing angles $< \theta_{cr}$, the electron scattering optical depth integrated from the outer radial boundary inwards is smaller than 1. There is no photosphere for these viewing angles. This is consistent with that found in previous works
(Curd \& Narayan 2019; Dai et al. 2018). Observers having small viewing angle can see directly down to the very inner region of the accretion flow. Observers having viewing angles larger than $\theta_{cr}$ are blocked by the optically thick outflow to see the region very close to the black hole.

Figure \ref{fig:Rsphere} shows the time evolution of the photosphere along three viewing angles of $60^\circ$ (green line), $75^\circ$ (blue line) and $90^\circ$ (black line). In model M6, for any time snapshot, the photosphere is largest at the viewing angle of $90^\circ$. This is because that the gas density is highest at $90^\circ$. In this model, the gas is injected into the computational domain at $45R_s (1.35\times 10^{13} {\rm cm}) -49R_s (1.47 \times 10^{13} {\rm cm})$. However, the photosphere increases with time and can be much larger than the injection radii. This is due to the fact that the outflows go outwards, which makes the photosphere increasing with time.  In model M6, initially, the photosphere is smaller than $10^{14} {\rm cm}$. It takes roughly 10 days for the photosphere to become larger than $10^{14} {\rm cm}$. The time interval of 10 days roughly equals to the time taken for the outflow to move from the injection radii to $10^{14} {\rm cm}$. We find that the outflow at $\theta > 60^\circ$ moves outwards with a velocity $\sim 0.003 c$. The time interval taken for the outflow to move to $10^{14} {\rm cm}$ is $10^{14} {\rm cm} / 0.003c \sim 10^6 {\rm s} \sim 11 {\rm days}$. For model M7, the gas is injected into the computational domain at $8R_s (2.4\times 10^{13} {\rm cm}) -12R_s (3.6 \times 10^{13} {\rm cm})$. It takes roughly 3-4 days for the photosphere to move to  $10^{14} {\rm cm}$. This time interval is also that taken by the outflow to move to $10^{14} {\rm cm}$. In model M7, the outflow at $\theta > 60^\circ$ moves outwards with a velocity $\sim 0.01 c$. The time interval needed for the outflow to move to $10^{14} {\rm cm}$ is $10^{14} {\rm cm} / 0.01c \sim 3.3 \times 10^5 {\rm s} \sim 3.8 {\rm days}$. For model M7, we find that after $\sim 100$ days, the emission radius slightly decreases with time. For model M6, we only simulate the TDEs evolving to 32 days. The reason for the shorter simulated evolution period in model M6 is as follows. The timestep of integration of the simulations is roughly proportional to $R_s/c$. The simulation integration timestep in model M6 is roughly 10 times smaller than that in model M7. Therefore, the simulation of M6 is significantly more time consuming than M7. We do not find the decrease phase of emission radius with time in model M6 due to the shorter simulated evolving period. Observationally, the optical/UV TDEs have an emission radius (photosphere) in the range of $ 10^{14} {\ \rm cm}$ to $10^{16} {\ \rm cm}$ (Hung et al. 2017; van Velzen et al. 2020; Gezari 2021). Generally, the emission radius/photosphere (after 3-10 days of gas injection) found in our simulations is consistent with observations. The simulations show that there is an initially quick increase period of the emission location. This period is 3-10 days for a black hole with mass of $10^6-10^7 M_\odot$ disrupting a solar type star. The future observations may find the quick increase phase of the emission radius. In model M7, the result that emission radius slightly decreases with time after $\sim 100$ days is also consistent with observations.

Figure \ref{fig:Trad} shows the time evolution of the radiation temperature at the photosphere along three viewing angles of $60^\circ$ (green line), $75^\circ$ (blue line) and $90^\circ$ (black line). The radiation temperature is calculated as $T_{\rm rad}= (E_{\rm rad}/a)^{1/4}$, with $E_{\rm rad}$ and $a$ being radiation energy density and the radiation constant, respectively. In model M6, initially the radiation temperature decreases quickly with time in the initial 10 days. The initial quick decrease phase of radiation temperature corresponds to the inital quick increase phase of the emission radius. After 10 days, the radiation temperature is in the range of $10^{4} {\rm K}-5\times 10^4 {\rm K}$ and nearly a constant with time. In model M7, the radiation temperature is also roughly a constant with time. Also, the radiation temperature is in the range of $10^{4} {\rm K}-6\times 10^4 {\rm K}$. Observations show that the black body radiation temperature is in the range of $10^{4} {\rm K}-5\times 10^4 {\rm K}$ (Hung et al. 2017; van Velzen et al. 2020; Gezari 2021). Also, observations show that the time variation of the radiation temperature is very small. Generally, the radiation temperature (after 3-10 days of gas injection) found in our simulations is consistent with observations. Also, our simulations show that the time variation of the radiation temperature is very small, which is also consistent with observations. Model M6 indicates that future observations may capture a quick decrease phase of radiation temperature, which last about 10 days.

We also calculate the optical/UV radiation luminosities as follows. In our models, we first find out the location of the photosphere. Second, if the radiation temperature at the photosphere is in the regime of $10^{4} {\rm K}-5\times 10^4 {\rm K}$, we calculate the radiation flux as $F_{\rm rad}=\sigma T_{\rm rad}^4$ (with $\sigma$ being the Stefan-Boltzman constant). Finally, to get the UV/optical luminosity, we integrate radiation flux over the photosphere for every viewing angles $\int F_{\rm rad} dS$, with $S$ being the surface area covering the photosphere. This method is not accurate but can roughly estimate the optical/UV radiation luminosity from the optically thick emitter. The reason is as follows. At the photosphere, the radiation is black body. Therefore, the optically thick photosphere mainly emits optical/UV photons with peak luminosity at the wavelength corresponding to the radiation temperature. We can compare the optical/UV luminosity calculated in our model to observations. Figure \ref{fig:Lum} shows the time evolution of the optical/UV luminosity calculated in our two models. It is clear that, after the photosphere becoming larger than $10^{14} {\rm cm}$, the luminosity is well in the observational range of $10^{43} - 10^{45} {\rm erg/s}$ (Hung et al. 2017; van Velzen et al. 2020; Gezari 2021).

\section{Summary and Discussions}
We perform radiation hydrodynamic simulations to study the optical/UV emission of the optical/UV TDEs. We assume that after disruption, the fall back bound debris can be quickly circularized.
Our numerical simulations take into account the specific conditions as described in the text for TDEs. The first one is that we inject gas at the theoretically predicted radii of the debris accretion disk. Second, the gas injection rate is set equal to the disrupted debris fallback rate. Our simulation results for the emission radius, the radiation temperature and the luminosity, as well as the evolution of these quantities with time, generated by reprocessing the emission of accretion flow in the optically outflows are consistent with observations in optical/UV TDEs. Our study provides a strong theoretical basis for understanding the origin of optical/UV TDEs, especially the evolution of emission radius, the radiation temperature and the luminosity in optical/UV TDEs, which are the key advancements in our simulations. Here we would like to mention that several other models have been proposed to explain the emission origin of optical/UV TDEs, such as Dai et al. (2018), Curd \& Narayan (2019) in the framework of reprocessing the emission of accretion flow in the optically thick outflows, and Piran et al. (2015), Jiang et al. (2016) in the shock collision scenarios, however both of which are focused on the stage around the peak luminosity of TDEs, and the evolution of the emission radius, the radiation temperature and the luminosity, are not strictly investigated.

The spectrum is not calculated in this paper. However, spectrum along different viewing angles is important to interpret observations. In future, we plan to calculate the spectrum of the circularized accretion flow system of TDEs by performing 3D Monto Carlo radiative transfer simulations.

In this paper, we inject gas at the circularization radius to mimic the fall back of debris. This is a simplification. In reality, when the debris falls back, it will penetrate the wind and then arrive at the circularization radius. When the debris penetrates winds, there may be shocks, which may affect the radiation properties of the system. In future, we plan to study more realistic circularized accretion flow in three-dimensions to take into account the self-consistent fall back of debris.

Our simulations are performed in pseudo-Newtonian potential. We also do not include magnetic field. We use an anomalous viscous stress to mimic the angular momentum transfer of the Maxwell stress. Due to the absence of both magnetic field and general relativistic potential, jet is not present in our simulations. Our models can not be applied to jetted TDEs. In future, it is quite necessary to perform general relativistic magnetohydrodynamic simulations taking into account the special conditions for TDEs.

\section*{Acknowledgments}
D. Bu thanks G. Mou for useful discussions. D. Bu is supported by the Natural Science Foundation of China (grants 12173065). E. Qiao is supported by the National Natural Science Foundation of China (grant 12173048) and NAOC Nebula Talents Program.  X. Yang is supported by the Natural Science Foundation of China (grant 11973018). This work made use of the High Performance Computing Resource in the Core Facility for Advanced Research Computing at Shanghai Astronomical Observatory.

\section*{Data availability}
The data underlying this article will be shared on reasonable request to the corresponding author.

\end{document}